\documentclass[aps,pre,amsmath,lengthcheck,superscriptaddress]{revtex4-2}

\usepackage{graphicx}\graphicspath{ {figures/} }
\usepackage{hyperref}
\hypersetup{colorlinks,allcolors=blue,breaklinks}

\newcommand{\be}{\begin{equation}}
\newcommand{\ee}{\end{equation}}
\newcommand{\ba}{\begin{align}}
\newcommand{\ea}{\end{align}}
\newcommand{\bi}{\begin{itemize}}
\newcommand{\ei}{\end{itemize}}

\newcommand{\tr}[1]{\mathrm{tr}\left\{#1\right\}}

\newcommand{\la}{\left\langle}
\newcommand{\ra}{\right\rangle}
\newcommand{\pd}{\partial}

\newcommand{\bla}{bla\\bla\\bla\\bla\\bla}

\newcommand{\mc}[1]{\mathcal{#1}}

\begin{document}

\title{Time-averaged quantum annealing for weak processes}

\author{Pierre Naz\'e}
\email{pierre.naze@unesp.br}

\affiliation{\it Departamento de F\'isica, Instituto de Geoci\^encias e Ci\^encias Exatas, Universidade Estadual Paulista ``J\'ulio de Mesquita Filho'', 13506-900, Rio Claro, SP, Brazil}

\date{\today}

\begin{abstract}

The quantum Ising chain has shortcuts to adiabaticity when operated with weak processes. However, when exactly do the non-equilibrium effects of the Kibble-Zurek mechanism, inherent to the system, appear in the optimal protocols in such a context? I propose here that such contrasting difference occurs due to the manner by which one measures the excitation spent energy of the system. Therefore, in this work, I made a qualitative analysis of a quantum annealing procedure of the time-averaged excess work, where the system acquires as a diverging decorrelation time the heuristic Kibble-Zurek mechanism relaxation time. Four important effects are then observed: the absence of shortcuts to adiabaticity, the pausing effect around the critical point in the optimal protocol when the Kibble-Zurek mechanism holds, the persistence of the time-averaged work to avoid slowly-varying regime even for large switching times, and diverging fluctuations of the time-averaged work. In the end, by comparing the excess and the time-averaged excess works, I conclude that this last one is not useful to measure the excitation spent energy in weak processes, although brings an intuition to what happens in the strong driving case.

\end{abstract}

\maketitle

\section{Introduction}
\label{sec:intro}

Quantum annealing is a procedure by which a quantum system performs a finite-time process in its ground state~\cite{deffner2019quantum}. The work produced by the non-equilibrium excitations, inherent to the process, is therefore null. Important applications, mainly on adiabatic quantum computing on avoiding informational errors~\cite{Chakrabarti2023,Hegde2022,King2022,Khezri2022,Morita2008,soriani2022assessing,Soriani2022}, can be applied with such an idea.

A paradigmatic model to study quantum annealing is the quantum Ising model~\cite{mbeng2020quantum}. Due to the existence of a second-order phase transition, its dynamics are very rich, being well-explained by the Kibble-Zurek mechanism~\cite{deffner2019quantum,yukalov2015realization,damski2005simplest,del2014universality,saito2007kibble}. This heuristic description is basically a breakdown of the adiabatic theorem when the system operates in the thermodynamic limit, having as a main consequence universal exponents in the work performed.

In recent years, an effort has been made in understanding such system in the context of weak drivings, in order to produce possible applications in quantum computing~\cite{naze2022,puebla2022linear,naze2023kibble}. In particular, it has been found the existence of universal shortcuts to adiabaticity if admissible optimal protocols are extended to distribution functions~\cite{naze2023universal2}. This however seems to be in stark contradiction with the expected idea of non-equilibrium excitations associated with the Kibble-Zurek mechanism. Most probably, such effects must appear when high orders than the linear one are accounted for in the work performed by the system.

Even though, can linear-response framework give some intuition of what happens to the optimal protocols when the Kibble-Zurek mechanism is present? The key word here is the waiting time associated with the system~\cite{naze2023universal2}, which depends essentially on the way the excitation energy is measured. In particular, the measurement made with time-averaged work will furnish a waiting time with the same diverging behavior of the heuristic timescale of the Kibble-Zurek mechanism~\cite{naze2023kibble}, in contrast with the null waiting time furnished by the measurement made with the conventional work~\cite{naze2023universal2}.

The objective of this work is to make a qualitative analysis of a quantum annealing procedure of the quantum Ising chain when it is measured by the time-averaged work. For doing this, I am going to use the near-optimal protocol recently discovered in the literature~\cite{naze2023performance}. Four characteristics will be observed: the absence of shortcuts to adiabaticity, the pausing effect around the critical point in the optimal protocol when the Kibble-Zurek mechanism holds, the persistence of the time-averaged work to avoid slowly-varying regime even for large switching times, and diverging fluctuations of the time-averaged work. In the end, by comparing the excess and the time-averaged excess works, I conclude that this last one is not useful to measure the excitation spent energy in weak processes, although brings an intuition to what happens in the strong driving case.

\section{Excess work}
\label{sec:lrt}

Consider a quantum system with a Hamiltonian $\mc{H}(\lambda(t))$, where $\lambda(t)$ is a time-dependent external parameter. Initially, this system is in contact with a heat bath of temperature $\beta\equiv {(k_B T)}^{-1}$, where $k_B$ is Boltzmann's constant. The system is then decoupled from the heat bath and, during a switching time $\tau$, the external parameter is changed from $\lambda_0$ to $\lambda_0+\delta\lambda$. The average work performed on the system during this process is
\be
W \equiv \int_0^\tau \la\pd_{\lambda}\mc{H}(t)\ra\dot{\lambda}(t)dt,
\label{eq:work}
\ee
where $\partial_\lambda$ is the partial derivative for $\lambda$ and the superscripted dot is the total time derivative. The generalized force $\la\pd_{\lambda}\mc{H}(t)\ra$ is calculated using the trace over the density matrix $\rho(t)$
\be
\la A(t)\ra =\tr{A\rho(t)}
\ee
where $A$ is some observable. The density matrix $\rho(t)$ evolves according to Liouville equation
\be
\dot{\rho} =\mc{L}\rho:= -\frac{1}{i\hbar}[\rho,\mc{H}],
\ee
where $\mc{L}$ is the Liouville operator, $[\cdot,\cdot]$ is the commutator and $\rho(0)=\rho_c$ is the initial canonical density matrix. Consider also that the external parameter can be expressed as
\be
\lambda(t) = \lambda_0+g(t)\delta\lambda,
\ee
where to satisfy the initial conditions of the external parameter, the protocol $g(t)$ must satisfy the following boundary conditions
\be
g(0)=0,\quad g(\tau)=1. 
\label{eq:bc}
\ee

Linear response theory aims to express the average of some observable until the first order of some perturbation considering how this perturbation affects the observable and the non-equilibrium density matrix \cite{kubo2012}. In our case, we consider that the parameter considerably does not change during the process, $|g(t)\delta\lambda/\lambda_0|\ll 1$, for all $t\in[0,\tau]$. Using the framework of linear-response theory, the generalized force can be approximated until the first-order as
\begin{equation}
\begin{split}
\la\pd_{\lambda}\mc{H}(t)\ra =&\, \la\pd_{\lambda}\mc{H}\ra_0+\delta\lambda\la\pd_{\lambda\lambda}^2\mc{H}\ra_0 g(t)\\
&-\delta\lambda\int_0^t \phi_0(t-t')g(t')dt',
\label{eq:genforce-resp}
\end{split}
\end{equation}
where the $\la\cdot\ra_0$ is the average over the initial canonical density matrix. The quantity $\phi_0(t)$ is the so-called response function, which can be conveniently expressed as the derivative of the relaxation function $\Psi_0(t)$
\be
\phi_0(t) = -\frac{d \Psi_0}{dt},
\label{eq:resprelax}
\ee 
where
\be
\Psi_0(t)=\beta\langle\partial_\lambda\mathcal{H}(t)\partial_\lambda\mathcal{H}(0)\rangle_0+\mathcal{C}
\ee
being the constant $\mathcal{C}$ calculated via the final value theorem \cite{kubo2012}. We define its decorrelation time as
\be
\tau_c = \int_0^\infty \frac{\Psi_0(t)}{\Psi_0(0)}dt.
\ee
In this manner, the generalized force, written in terms of the relaxation function, is
\begin{equation}
\begin{split}
\la\pd_{\lambda}\mc{H}(t)\ra =&\, \la\pd_{\lambda}\mc{H}\ra_0-\delta\lambda\widetilde{\Psi}_0 g(t)\\
&+\delta\lambda\int_0^t \Psi_0(t-t')\dot{g}(t')dt',
\label{eq:genforce-relax}
\end{split}
\end{equation}
where $\widetilde{\Psi}_0(t)\equiv \Psi_0(0)-\la\pd_{\lambda\lambda}^2\mc{H}\ra_0$. Combining Eqs. (\ref{eq:work}) and (\ref{eq:genforce-relax}), the average work performed at the linear response of the generalized force is
\begin{equation}
\begin{split}
W = &\, \delta\lambda\la\pd_{\lambda}\mc{H}\ra_0-\frac{\delta\lambda^2}{2}\widetilde{\Psi}_0\\
&+\delta\lambda^2 \int_0^\tau\int_0^t \Psi_0(t-t')\dot{g}(t')\dot{g}(t)dt'dt.
\label{eq:work2}
\end{split}
\end{equation}

We remark that in thermally isolated systems, the work is separated into two contributions: the quasistatic work $W_{\rm qs}$ and the excess work $W_{\rm ex}$. We observe that only the double integral on Eq.~(\ref{eq:work2}) has ``memory'' of the trajectory of $\lambda(t)$. Therefore the other terms are part of the contribution of the quasistatic work. Thus, we can split them as
\be
W_{\rm qs} = \delta\lambda\la\pd_{\lambda}\mc{H}\ra_0-\frac{\delta\lambda^2}{2}\widetilde{\Psi}_0,
\ee  
\begin{equation}
\begin{split}
W_{\text{ex}} = \delta\lambda^2 \int_0^\tau\int_0^t \Psi_0(t-t')\dot{g}(t')\dot{g}(t)dt'dt.
\label{eq:wirrder0}
\end{split}
\end{equation}
In particular, the excess work can be rewritten using the symmetry property of the relaxation function, $\Psi(t)=\Psi(-t)$ (see Ref.~\cite{kubo2012}),
\begin{equation}
\begin{split}
W_{\text{ex}} = \frac{\delta\lambda^2}{2} \int_0^\tau\int_0^\tau \Psi_0(t-t')\dot{g}(t')\dot{g}(t)dt'dt.
\label{eq:wirrder}
\end{split}
\end{equation}

We remark that such treatment can be applied to classic systems, by changing the operators to functions, and the commutator by the Poisson bracket \cite{kubo2012}.

\section{Time-averaged excess work}
\label{sec:taew}

Thermally isolated systems performing an adiabatic driven process can be interpreted as having a random decorrelation time \cite{naze2022adiabatic}. Therefore, at each instant of time that the process is performed, the relaxation function changes with it. This is very similar to what happens with systems performing an isothermal process, where the stochastic aspect of the dynamics changes the relaxation function. In this case, we take a stochastic average on the work to correct such an effect. In the case of thermally isolated systems, I propose as a solution the following time-averaging
\be
\overline{W}(\tau)=\frac{1}{\tau}\int_0^\tau W(t)dt.
\ee
Such quantity can be measured in the laboratory considering an average in the data set of processes executed in the following way: first, we choose a switching time $\tau$. After, we randomly choose an initial condition from the canonical ensemble and a time $t$ from a uniform distribution, where $0<t<\tau$. Removing the heat bath, we perform the work by changing the external parameter and collecting then its value at the end. The data set produced will furnish, on average, the time-averaged work.

In the following, I present how time-averaged work can be calculated using linear-response theory and how one can calculate the decorrelation time of the system. To do so, we define the idea of time-averaged excess work
\be
\overline{W}_{\rm ex} = \frac{1}{\tau}\int_0^\tau W_{\rm ex}(t)dt,
\ee
where $W=W_{\rm ex}+W_{\rm qs}$.

Now we observe how the time-averaged excess work can be calculated using linear-response theory. In Ref. \cite{naze2022}, I have shown that 
\be
\overline{W}_{\text{ex}}(\tau) = \delta\lambda^2\int_0^\tau\int_0^t \overline{\Psi}_0(t-t')\dot{g}(t)\dot{g}(t')dtdt',
\label{eq:TAexcesswork}
\ee
where
\be
\overline{\Psi}_0(t)=\frac{1}{t}\int_0^t \Psi_0(u)du,
\label{eq:relaxfuncaverage}
\ee
is the time-averaged relaxation function. This means that calculating the time-averaged excess work is the same as calculating the averaged excess work, but with a time-averaged relaxation function. Again, this is quite similar to what happens to systems performing isothermal processes, where a stochastic average is taken on the relaxation function. 

Now, when measured with time-averaged work, the thermally isolated system presents a decorrelation time. Indeed, the conditions such that linear-response theory is compatible with the Second Law of Thermodynamics are \cite{naze2020compatibility}
\be
\widetilde{\overline{\Psi}}_0(0)<\infty,\quad \Hat{\overline{\Psi}}_0(\omega)\ge 0,
\ee
where $\widetilde{\cdot}$ and $\hat{\cdot}$ are respectively the Laplace and Fourier transforms. Therefore, analogously to what happens in an isothermal process, we define a new decorrelation time
\be
\overline{\tau}_c := \int_0^\infty\frac{\overline{\Psi}_0(t)}{\overline{\Psi}_0(0)}dt=\frac{\widetilde{\overline{\Psi}}_0(0)}{\overline{\Psi}_0(0)}<\infty.
\label{eq:TArelaxtime}
\ee

\subsection{Optimization of time-averaged excess work}

Consider the time-averaged excess work rewritten in terms of the protocols $g(t)$ instead of its derivative
\begin{align}
    \overline{W}_{\rm ex} =& \frac{\delta\lambda^2}{2}\overline{\Psi}(0)+\delta\lambda^2\int_0^\tau \dot{\overline{\Psi}}_0(\tau-t)g(t)dt\\&-\frac{\delta\lambda^2}{2}\int_0^\tau\int_0^\tau \ddot{\overline{\Psi}}(t-t')g(t)g(t')dt dt'.
\end{align}
Using calculus of variations, we can derive the Euler-Lagrange equation that furnishes the optimal protocol $g^*(t)$ of the system that will minimize the time-averaged excess work \cite{naze2022optimal}
\be
\int_0^\tau \ddot{\Psi}_0(t-t')g^*(t')dt' = \dot{\Psi}_0(\tau-t).
\label{eq:eleq}
\ee
In particular, the optimal irreversible work will be \cite{naze2022optimal}
\be
\overline{W}_{\rm ex}^* = \frac{\delta\lambda^2}{2}\overline{\Psi}_0(0)+\frac{\delta\lambda^2}{2}\int_0^\tau \dot{\overline{\Psi}}_0(\tau-t)g^*(t)dt.
\ee
Also, the Euler-Lagrange equation \eqref{eq:eleq} furnishes also the optimal protocol that minimizes the variance of the time-averaged work \cite{naze2023optimal}. In this case, the optimal variance of the time-averaged work is
\be
\sigma_{\rm W}^{2^*} = \frac{\beta\delta\lambda^2}{4}\overline{\Psi}_0(0)+\frac{\beta\delta\lambda^2}{4}\int_0^\tau \dot{\overline{\Psi}}_0(\tau-t)g^*(t)dt.
\label{eq:optvariance}
\ee
In Ref.~\cite{naze2023universal}, the following universal solution was found
\be
g^*(t) = \frac{t+\overline{\tau}_w}{\tau+2\tau_w}+\sum_{n=0}^{\infty}\frac{a_n (\delta^{(n)}(t)-\delta^{(n)}(\tau-t))}{\tau+2\overline{\tau}_w},
\ee
where the time-averaged waiting time $\overline{\tau}_w$ is defined as
\be
\overline{\tau}_w = \widetilde{\overline{\Psi}}(0),
\ee
which is equal to the time-averaged decorrelation time for isothermal processes. In particular, one can consider as a near-optimal protocol only the continuous linear part, with an error around or less than $8\%$~\cite{naze2023performance}.

The objective of this work is to perform a qualitative quantum annealing in the quantum Ising model in the context of weak drivings but in its time-averaged version. By contrast with the shortcut to adiabaticity found in the context of weak drivings, here the optimal protocol will lead to non-zero time-averaged excess work. Therefore, we will observe what are the effects that the diverging time produces in the optimization of the protocol and work.

\section{Time-averaged quantum annealing}

In Ref.~\cite{naze2022}, my co-workers and I have shown that the relaxation function per number of spins for the transverse-field quantum Ising chain is
\be
\Psi_N(t)=\frac{16}{N}\sum_{n=1}^{N/2}\frac{J^2}{\epsilon^3(n)}\sin^2{\left(\left(\frac{2n-1}{N}\right)\pi\right)}\cos{\left(\frac{2\epsilon(n)}{\hbar}t\right)},
\ee
where
\be
\epsilon(n)=2\sqrt{J^2+\Gamma_0^2-2 J \Gamma_0 \cos{\left(\left(\frac{2n-1}{N}\right)\pi\right)}},
\ee
being $\Gamma_0$ the initial value of the magnetic field. The time-averaged relaxation function per number of spins will be
\be
\overline{\Psi}_N(t)=\frac{16}{N}\sum_{n=1}^{N/2}\frac{J^2}{\epsilon^3(n)}\sin^2{\left(\left(\frac{2n-1}{N}\right)\pi\right)}{\rm sinc}{\left(\frac{2\epsilon(n)}{\hbar}t\right)},
\label{eq:TApsi}
\ee
where
\be
{\rm sinc}(x) = \frac{\sin{(x)}}{x}.
\ee

Given the time-averaged relaxation function per number of spins~\eqref{eq:TApsi}, and using Eq.~\eqref{eq:TArelaxtime}, the time-averaged waiting time will be
\be
\overline{\tau}_w = \frac{\sum_{i=1}^{N/2}\frac{\pi\hbar}{\epsilon^4(n)}\sin^2{\left(\left(\frac{2n-1}{N}\right)\pi\right)}}{\sum_{i=1}^{N/2}\frac{4}{\epsilon^3(n)}\sin^2{\left(\left(\frac{2n-1}{N}\right)\pi\right)}},
\ee
which is naturally measured in units of $\hbar/J$. In Ref.~\cite{naze2023kibble} it is shown that it diverges accordingly to Kibble-Zurek mechanism prediction when $\Gamma_0\rightarrow J$.

\subsection{Near-optimal protocol}

\begin{figure}[h]
    \centering
    \includegraphics[scale=0.5]{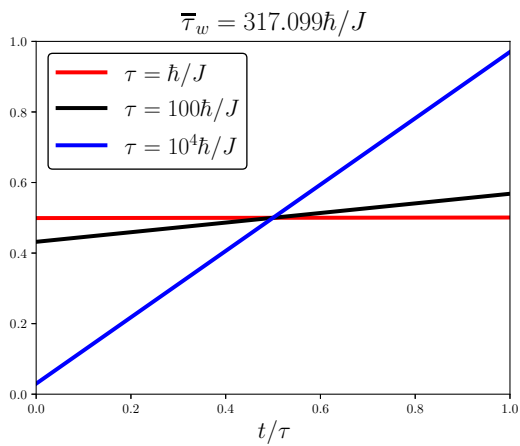}
    \caption{Optimal protocol for $\tau_w=317.099$ $\tau=1,100,10000$ (red, black and blue lines respectively), all measured in units of $\hbar/J$. For $\tau\ll\tau_w$, we are in the regime where Kibble-Zurek mechanism holds, and we observe the existence of pausing around the critical point. For $\tau\gg\tau_w$, the optimal protocol has the universal behavior of being a line connecting the initial and final point. It was used $\hbar=1,J=1,\Gamma_0=0.999995,\delta\Gamma=0.00001,N=10^4$.}
    \label{fig:optprot}
\end{figure}

To illustrate the effects of the diverging times in a set of parameters where linear-response theory holds~\cite{naze2022}, I choose the following parameters: $\hbar=1,J=1,\Gamma_0=0.999995,\delta\Gamma=0.00001,N=10^4$. In this case, the time-averaged waiting time will be $\overline{\tau}_w=317.099\hbar/J$. The continuous linear part of the optimal protocol, considering $\tau=1,100,10000$, in units of $\hbar/J$, is depicted in Fig.~\ref{fig:optprot}.

For switching times lesser than the time-averaged waiting time, the system is performing a quantum quench and presents the effects of Kibble-Zurek mechanism. Observe that, in this case, the effect of pausing around the critical point in the optimal protocol is similar to the sudden process case, where $\tau\rightarrow 0$~\cite{naze2022optimal}. However, here the switching time is much greater than 0, being the appearance of this effect only a manifestation of the ratio between the switching and diverging time-averaged waiting times. Observe also that such an effect is identical to what has been obtained in previous results in the literature~\cite{lidar2020why,passarelli2019improving,albash2021comparing}. For switching times greater than the time-averaged waiting time, the optimal protocol tends to be a line connecting the initial and final point, which is a universal behavior predicted by our works~\cite{naze2022optimal,naze2023universal}. This effect happens only for a finite number of particles, where one can find switching times that escape from Kibble-Zurek mechanism effects~\cite{naze2023kibble}. Finally, in the thermodynamic limit, ideally one can still find very small parameters where linear-response theory holds. In this case, however, there is no slowly-varying regime, and the optimal protocol should only appear with the pausing effect around the critical point.

\begin{figure}[h]
    \centering
    \includegraphics[scale=0.5]{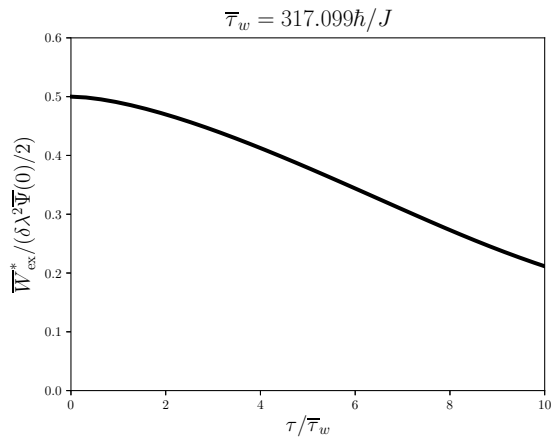}
    \caption{Near-optimal time-averaged excess work for quantum Ising model. Even for reasonable high values of $\tau$ in comparison to its time-averaged waiting time, the excess work is close to its maximum value. Also, the system presents a persistence to avoid the slowly-varying regime, which indicates that the non-equilibrium effects Kibble-Zurek are still in the system, even in its optimal version.}
    \label{fig:wex}
\end{figure}

\subsection{Near-optimal time-averaged excess work}

I now analyze the optimal work with the same parameters used in the previous section. The analysis will be qualitative since I am going to use only the continuous linear part, which gives a nice approximated result~\cite{naze2023performance}. Fig.~\ref{fig:wex} depicts the result. First, I observe that the optimal protocol is not a shortcut to adiabaticity, by contrast with its normal case~\cite{naze2023universal2}. Also, because of diverging time-averaged waiting time, the time-averaged excess work has values close to its maximum value even for reasonably high switching times. Indeed, the convergence to the slowly-varying regime is very slow, and even for switching times of $10$ times the time-averaged waiting time the excess work is not close to zero. This indicates that the non-equilibrium effects of the Kibble-Zurek mechanism persist in the system, even for large times and in its optimal version. Finally, in the thermodynamic limit, there is no slowly-varying regime, and the time-averaged excess work should only appear with its maximum value.

\subsection{Near-optimal fluctuations\\ of the time-averaged excess work}

The expression of the variance of the time-averaged work is given by Eq.~\eqref{eq:optvariance}, where we observe that it is proportional to $\beta$. Since the system ideally starts with $T=0$, then $\beta=\infty$. Also, the relaxation function does not depend on $\beta$, and therefore the optimal variance of the work diverges. Of course, this atypical scenario is just an ideal situation, and experiments performed at really low temperatures should furnish high but finite fluctuations. Also, the scenario in the thermodynamic limit does not change, since the value of the time-averaged excess work achieves its maximum value.

\subsection{Usefulness of time average work}

The consequences of measuring the excitation spent energy via the time-averaged work shows that it is not a good quantity in comparison to its conventional part. This is a conclusion for the weak processes scenario. However, this brings a perspective to strong driving scenarios as well. Indeed, since the waiting time of the quantum Ising model measured with excess work is zero, producing, therefore, shortcuts to adiabaticity, probably in higher orders the Kibble-Zurek mechanism relaxation time should appear. This will break down the complete suppression of excitation spent energy, as we have witnessed in the time-averaged case.

\section{Final remarks}

In this work, I made a qualitative analysis of the quantum annealing process of the quantum Ising chain in the scenario of measuring the excitation spent energy via the time-averaged work. Four important characteristics were observed: lack of shortcuts to adiabaticity, pausing effect around the critical point in the optimal protocol when Kibble-Zurek mechanism effects holds, persistence to avoid slowly-varying regime, and diverging fluctuations of the time-averaged excess work. Therefore, I conclude that this way of measuring the excitation spent energy is not useful. Even so, it brings a perspective of what would happen if higher orders were included in the  conventional work since the non-equilibrium effects of the Kibble-Zurek mechanism would probably appear.

\bibliography{TAQALR.bib}
\bibliographystyle{apsrev4-2}

\end{document}